\documentclass[useAMS,usenatbib]{mn2e}
\usepackage{graphicx,amssymb,color}
\usepackage[normalem]{ulem}

\title[No-planet WD rocky pollution]
{Orbit decay of 2-100 au planetary remnants around white dwarfs with no gravitational assistance from planets}

\author[]{Dimitri Veras$^{1,2,3}$\thanks{E-mail: d.veras@warwick.ac.uk}\thanks{STFC Ernest Rutherford Fellow},
Yusuf Birader$^{3}$, 
Uwais Zaman$^{3}$
\\
$^{1}$Centre for Exoplanets and Habitability, University of Warwick, Coventry CV4 7AL, UK
\\
$^{2}$Centre for Space Domain Awareness, University of Warwick, Coventry CV4 7AL, UK
\\
$^{3}$Department of Physics, University of Warwick, Coventry CV4 7AL, UK
}

\pubyear{2021}

\begin{document}
\label{firstpage}
\pagerange{\pageref{firstpage}--\pageref{lastpage}}
\maketitle

\begin{abstract}
A widely-held assumption is that each single white dwarf containing observable rocky debris requires the presence of at least one terrestrial or giant planet to have gravitationally perturbed the progenitor of the debris into the star. However, these planets could have been previously been engulfed by the star or escaped the system, leaving behind asteroids, boulders, cobbles, pebbles, sand and dust. These remaining small bodies could then persist throughout the host star's evolution into a white dwarf at $\approx 2-100$~au scales, and then be radiatively dragged into the white dwarf without the help of a planet. Here we identify the parameter space and cooling ages for which this one metal-pollution mechanism is feasible by, for the first time, coupling Poynting-Robertson drag, the Yarkovsky effect and the YORP effect solely from rapidly dimming white dwarf radiation. We find that this no-planet pollution scenario is efficient for remnant $10^{-5}-10^{-4}$~m dust up to about 80 au, $10^{-4}-10^{-3}$~m sand up to about 25 au and $10^{-3}-10^{-2}$~m small pebbles up to about 8 au, and perhaps $10^{-1}-10^{0}$~m small boulders up to tens of au. Further, young white dwarf radiation can spin up large strength-less boulders with radii $10^2-10^3$~m to destruction, breaking them down into smaller fragments which then can be dragged towards the white dwarf. Our work hence introduces a planet-less metal-pollution mechanism that may be active in some fraction of white dwarf planetary systems.
\end{abstract}


\section{Introduction}

White dwarf planetary systems are abundant. Dedicated surveys of single white dwarfs whose photospheres contain unambiguous signatures of rocky debris (so-called ``pollutants") suggest that between 25 and 50 per cent of these degenerate stars host planetary systems \citep{zucetal2003,zucetal2010,koeetal2014}. In terms of raw numbers, the tally is well over 1,000 \citep{couetal2019}, and is set to jump by over an order of magnitude with recent {\it Gaia} discoveries \citep{genetal2019}.   

However, the number of white dwarf systems for which actual planets (mostly round objects with radii above $10^3$ km) have been discovered is only a handful \citep{thoetal1993,sigetal2003,luhetal2011,ganetal2019,vanetal2020,blaetal2021}. This small value results from both the frequency of planets which do exist around white dwarfs, and the considerable observational challenges of discovering such planets \citep[e.g.][]{sanetal2016,marsh2018,vanvan2018,corkip2019}. Although discoveries are mounting and the statistics will improve significantly \citep{peretal2014,danetal2019,tamdan2019,kanetal2021}, there exists a signficiant gap in detection numbers between planets and pollution detections due to observational bias.

Consequently, understanding how dependent pollution is on the presence of planets is important. Before the discovery of any white dwarf planets, \cite{graetal1990} speculated that dusty signatures around G29-38 could have arisen from a catastrophic event involving asteroids during the white dwarf phase, followed by inward radiative drag due to the Poynting-Robertson effect. Subsequently, the idea that dusty debris is formed by a tidal destruction event has become canonical \citep{jura2003,jura2008}, spurred on by evidence that, in at least some systems, tidal disruption is ongoing \citep{vanetal2015,vanderboschetal2020,faretal2021,guietal2021}.

The tidal disruption of asteroids requires the presence of planets because asteroids are too large to be radiatively dragged via Poynting-Robertson drag from a distance of a few au to the white dwarf Roche sphere: a few au represents the minimum distance at which a primordial asteroid can avoid engulfment from the progenitor giant branch star \citep{musvil2012,adablo2013}. Hence, planets are needed to gravitationally perturb the asteroids towards the white dwarf. \cite{broetal2021} recently set out a detailed road map for pollution in white dwarf systems under this assumption.

Challenges to this canonical theory of the dynamical origin of pollution have arisen from the alternate ideas of second-generation formation \citep{perets2011,perkey2013,beasok2014,schdre2014,voletal2014,hogetal2018}, evaporation of planetary atmospheres \citep{schetal2019}, and impact ejecta from collisions with planets \citep{verkur2020}. However, second-generation formation around single white dwarfs is feasible only for high disc masses ($> 10^{23}$ kg) \citep{vanetal2018}, and the other two ideas still require the presence of planets.

Here, we consider systems where planets have already been removed through either engulfment into the star or escape due to gravitational scattering. For example, within protoplanetary discs, fast migration could drive planets into the star \citep[e.g.][]{ogietal2015,parjoh2018,raymor2020}. During the main sequence, tidal inspiral of close-in planets will eventually lead to destruction \citep[e.g.][]{hamsch2019,alvetal2021,maful2021}, while exterior planets could escape through scattering \citep[e.g.][]{davetal2014,puwu2015,petetal2020,wanetal2020}. During the giant branch phases, the increasing size of the star engulfs planets out to au-scale distances \citep[e.g.][]{musvil2012,norspi2013,madetal2016,ronetal2020}. During the white dwarf phase, escape \citep[e.g.][]{maletal2020a,maletal2020b,maletal2021} tidal inspiral \citep{veretal2019b,verful2019} and atmospheric evaporation \citep{schetal2019} could all remove planets.

\begin{figure}
\includegraphics[width=8.0cm]{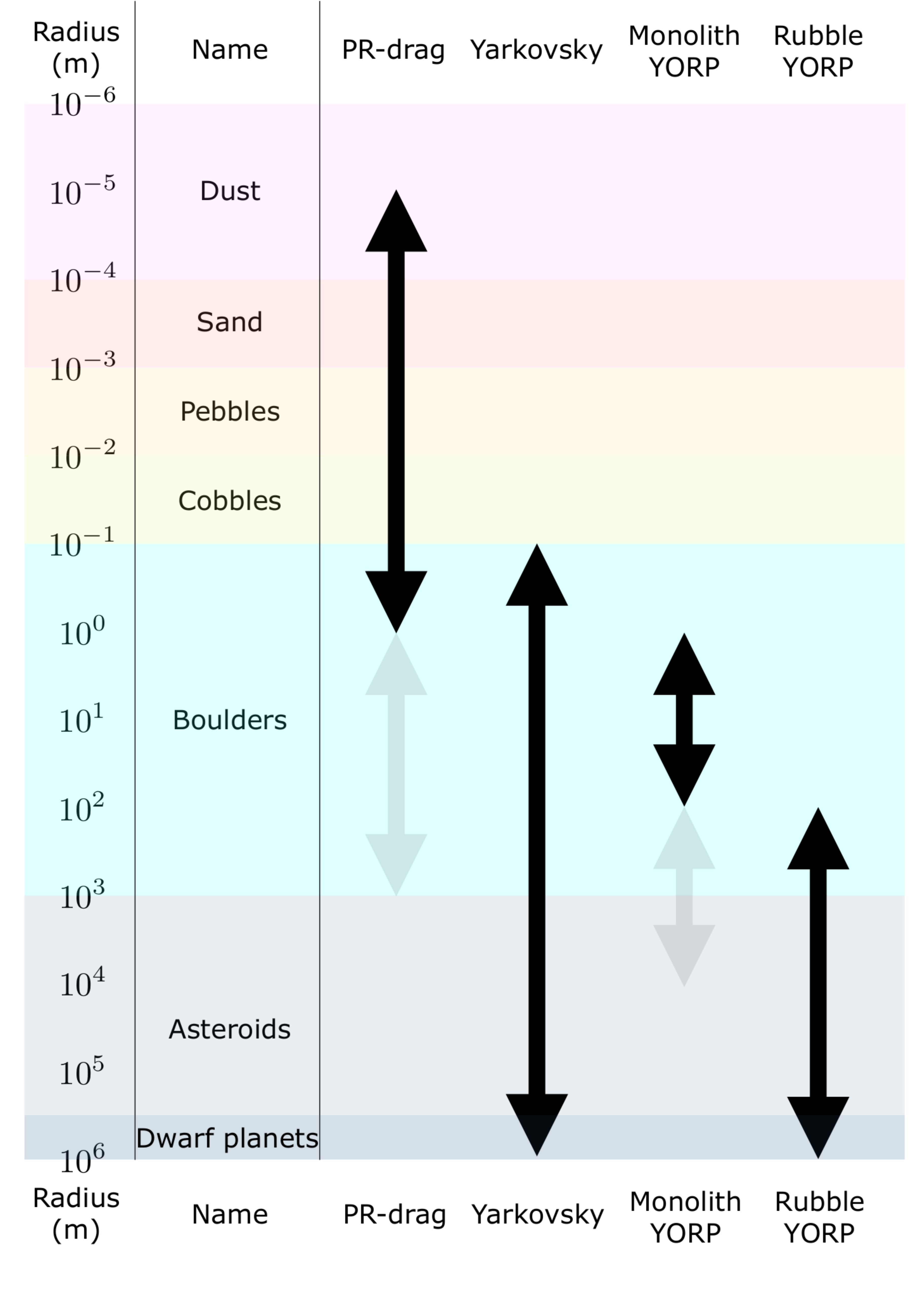}
\caption{
Size definitions of objects used throughout this paper, along with the size ranges of applicability of different radiative forces. Translucent arrows indicate feasible but unlikely applicability. 
}
\label{Scales}
\end{figure}

Left behind could be fully-formed asteroids and smaller objects. As these objects are exposed to the star's changing mass and luminosity, disc or belt structures evolve according to a variety of collisional and radiative processes \citep{bonwya2010,donetal2010}, whereas individual boulders or asteroids may be subjected to radiative spin-up and drift from the Yarkovsky and YORP effects \citep{veretal2014b,veretal2015a,veretal2019a,versch2020}. The final system configuration at the birth of the white dwarf depends strongly on the fine details of these processes.

Our study starts at this epoch. For the first time, we quantify if and how the rapidly weakening radiation from a white dwarf alone can generate pollutants on arbitrary au-scale orbits left over from giant branch evolution by coupling Poynting-Robertson drag, the Yarkovsky effect and the YORP effect. Previous relevant work on au-scale radiative drag in white dwarf systems \citep{veretal2015b} assumed the presence of at least one planet. Consequently, their parameter space exploration was limited to initially highly eccentric debris orbits. Further, they did not incorporate the Yarkovsky and YORP effects in their calculations, and they explored only a four-order-of-magnitude size range of objects affected by Poynting-Robertson drag alone.

In this paper,
we first present definitions and equations in Section 2 before describing our simulations and results in Section 3.  Sections 4 and 5 then represent discussion sections. In Section 4, we discuss the potential size distributions and distance ranges of leftover planetary debris from the giant branch phases, and in Section 5 we discuss  potentially observable features which may be used to distinguish no-planet pollution from planet-rich pollution. We conclude in Section 6.

\section{Definitions and equations}

In this work, we investigate the evolution of objects which are smaller than planets. The term planet is defined as a mostly round object of radius greater than $10^3$ km. We further define seven types of smaller objects (dwarf planets, asteroids, boulders, cobbles, pebbles, sand, dust) and shoehorn their size ranges into the convenient order-of-magnitude scaling of Fig. \ref{Scales}. These objects are also considered to be primarily but not entirely round so that (i) a mean radius $R$ can be defined, (ii) the mass, mean radius and density $\rho$ (assumed constant) are simply related through a spherical approximation, (iii) and the YORP effect can be applied, which requires at least slight asymmetries. 

The physical processes that we consider encompass almost the entire range on the scale in Fig. \ref{Scales}. We will now go through each of these processes in turn. 

\subsection{Poynting-Robertson drag}

The first is Poynting-Robertson (PR) drag. In white dwarf planetary science, PR drag is usually invoked only within the immediate vicinity (a couple Solar radii) of the white dwarf \citep{rafikov2011a,rafikov2011b} or for highly eccentric orbits produced due to tidal disruption \citep{veretal2015b,nixetal2020,broetal2021,lietal2021,malamudetal2021}. Here, however, we apply PR drag on 
$10^{-3}-10^{2}$ au-scale distances of unperturbed objects of nearly any eccentricity. 

PR drag creates an inward drift, and changes the semimajor axis $a$ and eccentricity $e$ of an object secularly (i.e. averaged over anomalies or longitudes) according to \citep{wyawhi1950}

\begin{equation}
\left( \frac{da}{dt} \right)_{\rm PR} = -\left(\frac{1}{c^2}\right)
\frac{3\left(Q_{\rm ABS} + Q_{\rm REF}\right)\left(2 + 3e^2\right)L_{\star}}
{16 \pi R \rho a \left(1 - e^2\right)^{3/2}},
\label{PRa}
\end{equation}

\begin{equation}
\left( \frac{de}{dt} \right)_{\rm PR} = -\left(\frac{1}{c^2}\right)
\frac{15\left(Q_{\rm ABS} + Q_{\rm REF}\right)e L_{\star}}
{32 \pi R \rho a^2 \sqrt{1 - e^2}}.
\label{PRe}
\end{equation}

\noindent{}In Eqs. (\ref{PRa}-\ref{PRe}), $c$ is the speed of light, $L_{\star}$ is the time-dependent luminosity of the star, $Q_{\rm ABS}$ is a constant that describes the object's reflection efficiency, and $Q_{\rm REF}$ is a constant that describes the object's absorption efficiency. Hence, the drift is always inward, but is modulated by the sum of $Q_{\rm ABS} + Q_{\rm REF}$.

In this study, our goal is to place limits on the motion by establishing bounds. Hence, subsequently we assume $Q_{\rm ABS} + Q_{\rm REF} = 1$ and consider PR drag only in the object size regime of $R=10^{-5}-10^{0}$ m. Smaller objects, such as micron-sized dust, are subject to complex scattering properties from the Mie scattering regime, and may not be adequately modelled with the above equations. For boulders larger than $R=10^{0}$ m, the effect is applicable but negligible, and need not be considered (indicated by a translucent arrow on Fig. \ref{Scales}).  

\subsection{Yarkovsky effect}

The Yarkovsky effect can overwhelm PR drag -- by a factor of $(c/v)$ -- where $v$ is velocity. However, the Yarkovsky effect is a more complex function of the shape of the object, as well as its spin angular momentum and orbital angular momentum \citep{broz2006}. \cite{veretal2015a} attempted to combine these last two components in a self-consistent analytical way in terms of orbital elements by assuming a spherical object. They derived secular expressions for those elements. 

By considering only the maximum possible Yarkovsky drift, we can apply these secular expressions here. \footnote{The way to achieve these limiting values is to adopt, in the notation of \cite{veretal2015a}, constant values of $\mathcal{Q}_{11} = \mathcal{Q}_{22} = \mathcal{Q}_{33} =~1, \mathcal{Q}_{12} = 1/4, i=0^{\circ}, \Omega=0^{\circ}$, and $\omega=0^{\circ}$.} The maximum semimajor axis change due to the Yarkovsky effect equals

\[
\left( \frac{da}{dt} \right)_{\rm Yar, Max} = -\left(\frac{1}{c}\right)
\frac{3L_{\star}}
{64 \pi R \rho \sqrt{G M_{\star} a} \left(1 - e^2\right)}
\]

\begin{equation}
\ \ \ \ \ \ \ \ \ \ \ \ \ \ \ \ \ \ 
+ \mathcal{O}\left(\frac{1}{c^2}\right)
,
\label{Yarka}
\end{equation}

\noindent{}where $M_{\star}$ is the mass of the star and the $\mathcal{O}$ term represents a single term rather than a longer series. Computing the maximum eccentricity change due to the Yarkovsky effect is slightly more involved but equals the following compact expression:

\[
\left( \frac{de}{dt} \right)_{\rm Yar, Max} = -\left(\frac{1}{c}\right)
\frac{3L_{\star}}
{128 \pi R \rho \sqrt{G M_{\star} a^3}}
\]

\[
\ \ \ \ \ \ \ \ \ \ \ \ \ \ \ \ \ \ \ \ \ 
\times \left[
\frac
{e^2- 2\left(1-e^2\right)\left(1 - \sqrt{1 - e^2}\right)}
{e^3}
\right]
\]

\begin{equation}
\ \ \ \ \ \ \ \ \ \ \ \ \ \ \ \ \ \ 
+ \mathcal{O}\left(\frac{1}{c^2}\right)
.
\label{Yarke}
\end{equation}

\noindent{}Note that the term in square brackets is always positive or zero, and as $e \rightarrow 0$, $de/dt \rightarrow 0$.

Unlike PR drag, the Yarkovsky effect can shift objects inwards or outwards, repeatedly changing direction throughout the evolution. In order to roughly quantify this variability, we introduce a constant efficiency factor $\xi \in \left[-1,1\right]$ such that 

\begin{equation}
\left( \frac{da}{dt} \right)_{\rm Yar} = \xi\left( \frac{da}{dt} \right)_{\rm Yar, Max}
,
\end{equation}

\begin{equation}
\left( \frac{de}{dt} \right)_{\rm Yar} = \xi\left( \frac{de}{dt} \right)_{\rm Yar, Max}
.
\end{equation}

\noindent{}We are uninterested in the case of net outward migration ($\xi <0$), and in most instances adopt the maximum inward drift scenario ($\xi = 1$). 

Comparing Eqs. (\ref{PRa}-\ref{Yarke}) with one another illustrates that when both the Yarkovsky effect and PR drag operate simultaneously, then the Yarkovsky effect takes over. Hence, understanding when the Yarkovsky effect is ``activated" is important. This activation could occur only if the object is large enough to internally redistribute heat and emit it anisotropically, a capability which is also dependent on the object's spin, thermal conductivity and thermal capacity. We adopt the activation size range of $R=10^{-1}-10^{6}$~m that is given in Fig. \ref{Scales} based on the findings of \cite{veretal2015a}. The lower end of this range is uncertain, but, as we will show, does affect our results. Hence, Fig. \ref{Scales} indicates that each of PR drag and the Yarkovsky effect may dominate in the size range $R=10^{-1}-10^{0}$ m.

\begin{figure}
\includegraphics[width=8.5cm]{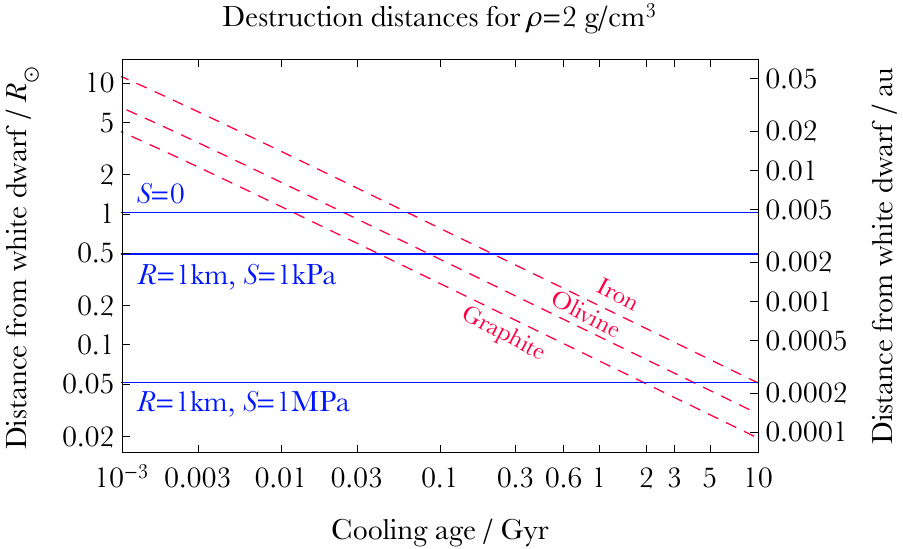}
\caption{
Determining whether an object first sublimates or disrupts upon drifting close to the white dwarf, as a function of the star's cooling age. The red dashed lines indicate sublimation distances for three different materials, and the three blue horizontal lines indicate different Roche radii. The plot suggests that all destruction activity occurs within a distance of about $10R_{\odot}$. After about 100 Myr, this value is decreased to about $1R_{\odot}$. For cooling ages of several Gyr, strong objects can easily reach within $0.2R_{\odot}$ before being destroyed.
}
\label{DesPlot}
\end{figure}

\begin{figure*}
\centerline{\LARGE \underline{Planet-less system pollutants: time evolution}}
\centerline{}
\centerline{
\includegraphics[width=8.5cm]{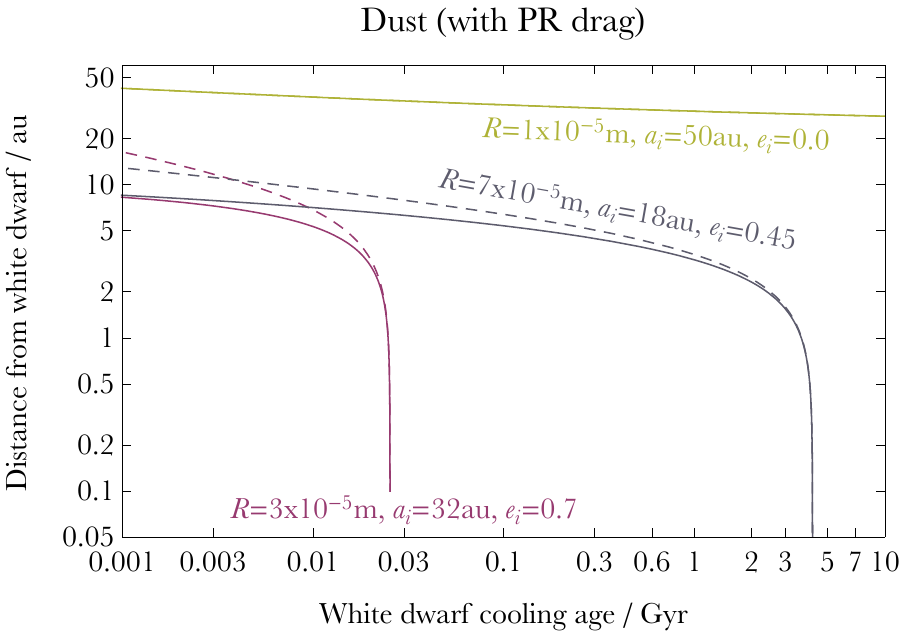}
\includegraphics[width=8.5cm]{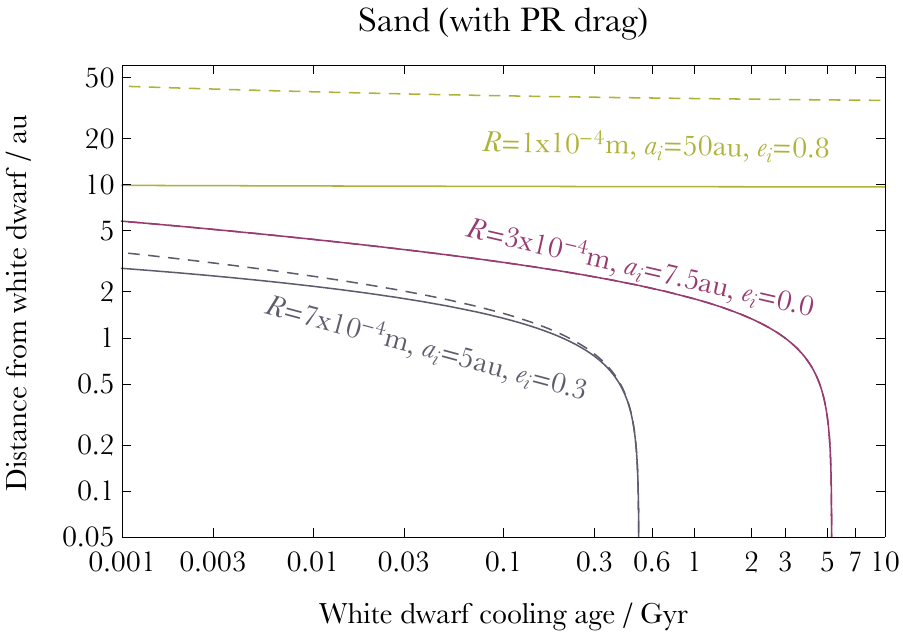}
}
\centerline{}
\centerline{
\includegraphics[width=8.5cm]{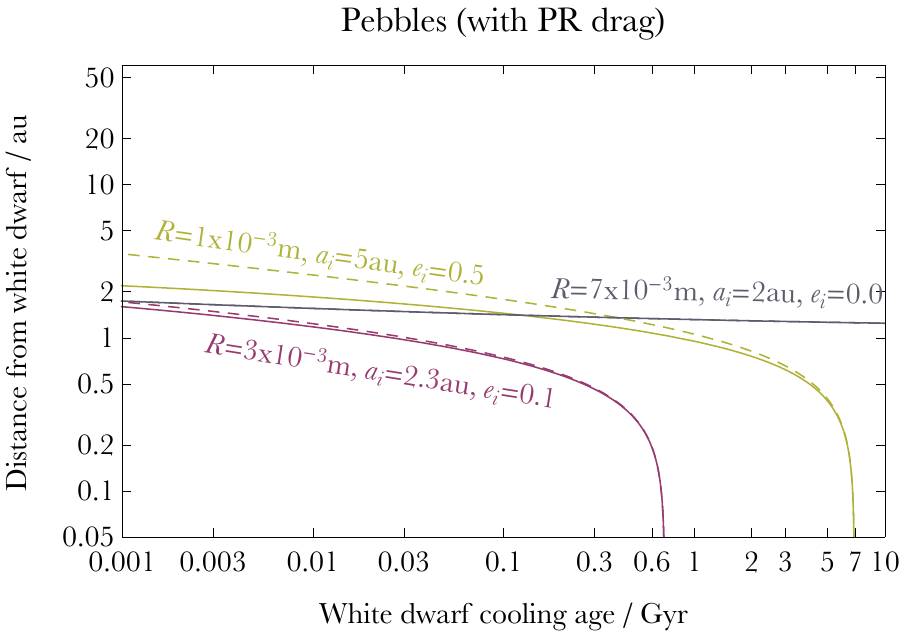}
\includegraphics[width=8.5cm]{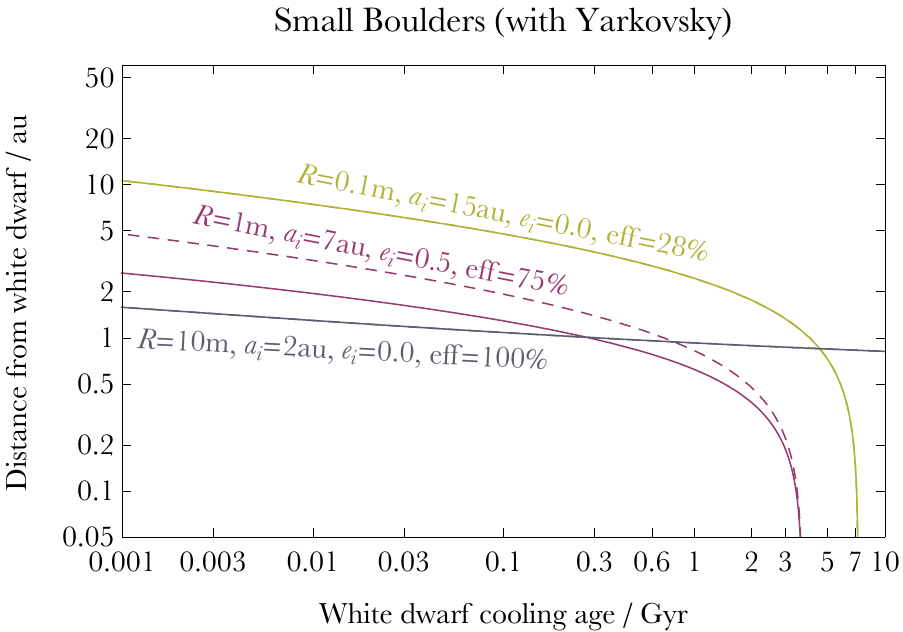}
}
\caption{
The time evolution of four types of objects (dust, sand, pebbles, small boulders) due to PR drag or the Yarkovsky effect. Within each pair of curves, the dashed one represents the semimajor axis, and the solid one represents the orbital pericentre. By tweaking the object radius, initial semimajor axis, initial eccentricity and Yarkovsky efficiency, one can generate accretion onto the white dwarf for any of these objects at any cooling age. Isolated cobbles ($R=10^{-2}-10^{-1}$~m), large boulders ($R=10^{1}-10^{3}$~m) and asteroids and dwarf planets ($R=10^{3}-10^{6}$~m) fail to accrete and will remain exterior to the star throughout its cooling.
}
\label{time}
\end{figure*}

\subsection{YORP effect}

Unlike PR drag and the Yarkovsky effect, the YORP effect does not alter orbital parameters. Instead, the YORP effect spins up and/or down the objects \citep{voketal2015}. The consequences are important for our study because if the object is spun up to break-up speed, then the resulting fragments could be small enough to be efficiently radiatively dragged into the white dwarf.

Like the Yarkovsky effect, the YORP effect is a nontrivial function of shape and internal structure, and can change spin direction throughout the evolution. A rough but useful maximum estimate for how quickly the object's spin rate, $\omega$, can change on secular timescales is \citep{scheeres2007,scheeres2018}

\begin{equation}
\left( \frac{d\omega}{dt} \right)_{\rm YORP,max}
=
\frac{3 \Phi}{4 \pi}
\left(\frac{\chi}{\rho R^2}\right)
\left[ \frac{1}{a^2\sqrt{1-e^2}} \frac{L_{\star}}{L_{\odot}} \right]
.
\label{YORP}
\end{equation}

This estimate represents a maximum estimate because Eq. (\ref{YORP}) assumes monotonic spin-up. In the equation,  the un-delimited coefficients represent constants which are fixed across all simulations, with $\Phi = 1 \times 10^{17}$\,kg\,m\,s$^{-2}$ being the Solar radiation constant. The terms in parenthesis are constants throughout an individual simulation, but may be altered across different simulations. In this sense, these three parameters ($\chi$,$\rho$,$R$) together represent a scaling factor for the results. Within this scaling factor, the term $\chi$ is a catch-all parameter which characterizes the asymmetry and obliquity of the object. A typical value which we will adopt is $\chi = 10^{-3}$. For spherical objects ($\chi = 0$), the YORP effect turns off. In this sense, we are deliberately being slightly inconsistent in our spherical definition of $R$ when $\chi=10^{-3}$.

Finally, the term in square brackets is time-variable throughout each simulation. Because Eqs. (\ref{PRa}-\ref{YORP}) all depend on $a$ and $e$, all those equations must be solved simultaneously. However, none of them can be solved without first detailing the time dependence of the stellar luminosity ($L_{\star}$), which we do in the next subsection.

\begin{figure*}
\centerline{\LARGE \underline{Planet-less system pollutants: close to white dwarf}}
\centerline{}
\centerline{
\includegraphics[width=8.5cm]{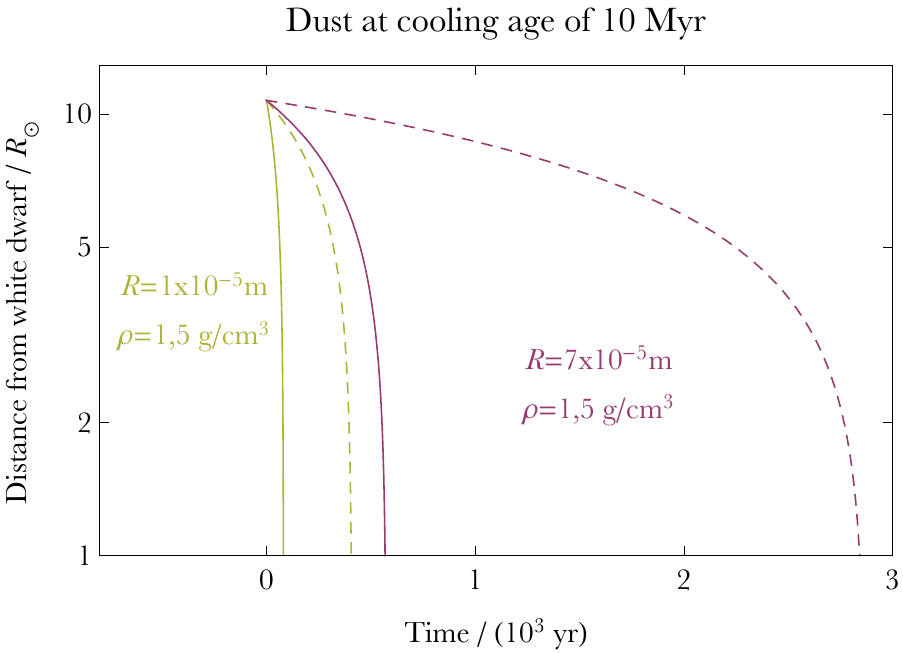}
\includegraphics[width=8.5cm]{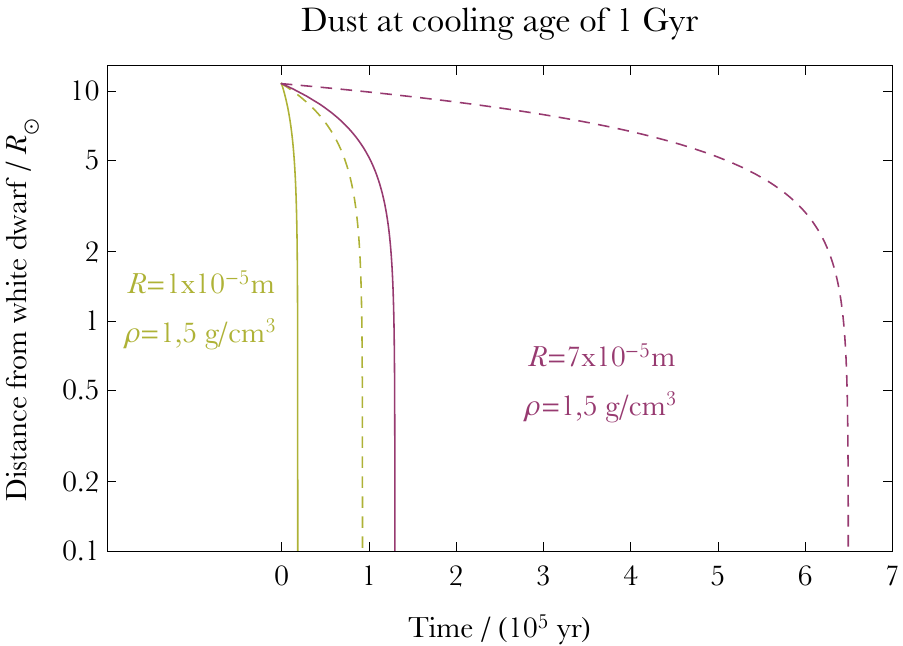}
}
\caption{
Behaviour of dust within $10R_{\odot}$ of the white dwarf. The particles are started on circular orbits; the solid lines represent the $\rho = 1$ g/cm$^3$ cases, and the dashed lines represent the $\rho = 5$ g/cm$^3$ cases. Time is measured from cooling ages of 10 Myr (left panel) and 1 Gyr (right panel).
}
\label{cool}
\end{figure*}

First, we discuss the size range of applicability of the YORP effect. Key observational evidence in the form of the spin and size characteristics of solar system boulders and asteroids guides our choices. Figure 2 of \cite{huetal2021} reveals a sharp natural line in this parameter space for $R=10^{2}-10^{6}$ m. In this size regime, almost no boulders or asteroids harbour a spin period shorter than about 2.3 hours. The likely reason is that 2.3 hours represents the ``spin barrier" for rubble pile asteroids with no material strength, such that any spin period shorter than 2.3 hours would break apart the asteroid. 

In contrast, boulders with $R=10^{0}-10^{2}$ m appear to easily bypass this spin barrier, with spin periods much shorter than 2.3 hours. These boulders probably then have material cohesive strength, and are more likely to represent ``monoliths" rather than rubble piles. The arrows on Fig. \ref{Scales} reflect this division in size. The translucent arrow indicates that monoliths larger than $10^2$ m do exist, but are rare. Monoliths as large as $R=10^4$ m are physically feasible \citep{zhaetal2021}.

The rotational break-up speed of an approximately spherical asteroid can be approximated as \citep{sansch2014,scheeres2018}

\begin{equation}
\omega_{\rm breakup} = 
\sqrt{
\frac{4\pi G \rho}{3} + \frac{4 S}{3 \rho R^2}
}
,
\label{breakup}
\end{equation}
 
\noindent{}where $S$ represents the object's material cohesive strength. The 2.3 hour spin barrier is approximately obtained in the rubble pile limit, by setting $S = 0$ Pa. Finally, note that upon reaching the white dwarf phase, an object's initial spin may be any value from $-\omega_{\rm breakup}$ to $0$ to $\omega_{\rm breakup}$.

\subsection{Luminosity}

A key parameter underpinning the manner of the object's evolution is the stellar luminosity, $L_{\star}$. A useful approximate analytical expression for a white dwarf's luminosity variation as a function of time is \citep{mestel1952}

\begin{equation}
L_{\star} = 3.26 L_{\odot} \left( \frac{M_{\star}}{0.6 M_{\odot}} \right)
\left(0.1 + \frac{t}{\rm Myr}  \right)^{-1.18}
.
\label{lum}
\end{equation}

This equation illustrates that the moment a white dwarf is born, its luminosity is $\sim 10L_{\odot}$. After just 1 Myr, its luminosity decreases to $\sim 1L_{\odot}$, whereas after 10 Myr, its luminosity decreases to $\sim 0.1L_{\odot}$. Hence, for the vast majority of a white dwarf's existence, $L_{\star} < L_{\odot}$, meaning that PR drag, the Yarkovsky effect and the YORP effect are generally less effectual than along the main sequence and especially along the giant branches.

\subsection{Sublimation and tidal destruction}

This time-varying luminosity can be used to determine the distance at which objects would sublimate \citep[e.g.][]{steetal2021}. Objects that are dragged towards the white dwarf will eventually be destroyed, either through sublimation, or by entering the disruption, or Roche sphere. Which of these two possible fates will occur is determined by the age of the white dwarf and the material properties of the object.

An approximate expression for the distance from the white dwarf at which a particular material will sublimate, $r_{\rm sub}$, is \citep{rafikov2011b}

\begin{equation}
r_{\rm sub} = \frac{R_{\star}}{2} \left(\frac{T_{\star}}{T_{\rm sub}}\right)^2,
\label{subT}
\end{equation}

\noindent{}where $R_{\star}$ is the radius of the white dwarf, $T_{\star}$ is the effective temperature of the white dwarf, and $T_{\rm sub}$ is the sublimation temperature of the object. We can combine Eq. (\ref{subT}) with Eq. (\ref{lum}) and with the equation for a blackbody

\begin{equation}
L_{\star} = 4\pi R_{\star}^2 \sigma T_{\star}^4,
\end{equation}

\noindent{}where $\sigma$ is the Stefan-Boltzmann constant, to obtain

\begin{equation}
r_{\rm sub} = \frac{1}{4}
\sqrt{ 
\frac{3.26 L_{\odot}}{\pi \sigma T_{\rm sub}^4}
\left( \frac{M_{\star}}{0.6 M_{\odot}} \right)
\left(0.1 + \frac{t}{\rm Myr}  \right)^{-1.18}
}
.
\label{subTfinal}
\end{equation}

In Fig. \ref{DesPlot}, we plot Eq. (\ref{subTfinal}) as a function of the cooling age of the white dwarf, which is defined as the time since the star became a white dwarf. We consider three materials \citep{rafgar2012}: iron ($T_{\rm sub} \approx 1600$ K), olivine ($T_{\rm sub} \approx 2100$ K) and graphite ($T_{\rm sub} \approx 2600$ K). The plot illustrates that for all cooling ages up to 10 Gyr, the sublimation distance ranges from about $0.02R_{\odot}$ to $10R_{\odot}$.

Now we compare these values to those of the white dwarf Roche sphere. The size of the Roche sphere depends on the spin, shape and internal strength of the object. We can roughly approximate the Roche sphere radius, $r_{\rm Roche}$, by a single formula that includes material cohesive strength as \citep{holmic2008,zhaetal2021}

\begin{equation}
r_{\rm Roche} = 
\left[
\left( \frac{8\pi}{15} + \frac{2\sqrt{3}S}{G R^2 \rho^2} \right)
\left( \frac{\rho}{M_{\star}} \right)
\right]^{-1/3}
,
\label{roche}
\end{equation}

\noindent{}where we have assumed a friction angle of $30^{\circ}$. Although Eq. (\ref{roche}) could underestimate $r_{\rm Roche}$ in the rubble pile limit ($S=0$ Pa), this underestimation may be compensated for by a change in the assumed spin of the object.

We plot Eq. (\ref{roche}) together with Eq. (\ref{subTfinal}) in Fig. \ref{DesPlot}, assuming $\rho = 2$ g/cm$^3$. Unlike the sublimation distance, the Roche radius is not time-dependent. The rubble pile Roche radius for this density is approximately $1R_{\odot}$, which is a commonly-used approximation in white dwarf planetary science. However, note the dependencies on physical parameters: decreasing the density by half would increase $r_{\rm Roche}$ by about 26 per cent. Further, as the material cohesive strength is increased, $r_{\rm Roche}$ decreases, potentially to within $0.05R_{\odot}$. Particularly notable is that strong objects, such as the theorized monolithic asteroid orbiting white dwarf SDSS J1288+1040 \citep{manetal2019}, can drift within $0.5R_{\odot}$ and survive.

\section{Simulations}

Having defined our terms and presented our equations of motion, we now proceed to describe our simulations. Throughout all simulations, we define the initial values of $a$ and $e$ as $a_{\rm i}$ and $e_{\rm i}$ (where ``initial" refers to the start of the white dwarf phase), and impose the condition $a_{\rm i}\left(1 - e_{\rm i}\right) >2$~au. This minimum initial separation roughly corresponds to the maximum giant branch engulfment distance of a typical white dwarf planetary system host. We hence also set $M_{\star} = 0.6M_{\odot}$, which represents a fiducial white dwarf mass value \citep{treetal2016,cumetal2018,elbetal2018,mccetal2020,barcha2021}. All of our simulations are run for 10 Gyr of white dwarf cooling, or until the object reaches a distance of $0.05R_{\odot}$.

\subsection{Overall time evolution}

We start by displaying representative time evolutions of objects in Fig. \ref{time}. These plots alone demonstrate that planets are not required for white dwarf pollution; smaller objects can achieve the feat by themselves. We chose only dust, sand, pebbles and small boulders because the condition $a_{\rm i}\left(1 - e_{\rm i}\right) >2$ au allows neither cobbles to be dragged close to the white dwarf through PR drag, nor asteroids to be sufficiently dragged there through the Yarkosvky effect. Hence, if one assumes no collisional evolution amongst the objects, then PR drag and the Yarkovsky effect may create a break in the size distribution of objects that could reach the white dwarf without the help of a planet.

The parameter choices for the example evolutions in the figure were chosen to showcase a variety of behaviours and the interplay between the dependencies on $R$, $a_{\rm i}$ and $e_{\rm i}$. Some evolutions do not end with the object reaching the immediate vicinity of the white dwarf, and others show accretions at a wide range of cooling ages. The figure also illustrates that the time evolution can be strongly dependent on the object's initial eccentricity, and even for high eccentricities (upper curves for sand), there is no guarantee that accretion will occur with 10 Gyr. The variety of evolutions in the figure is greater than what is computed due to PR drag in the high-eccentricity tidal disruption scenario with $e_{\rm i} > 0.99$ \citep{veretal2015b}.

The pebble plot illustrates that pebbles can persist at au-distances for several Gyr before being accreted. Because these objects are so large compared to dust or sand, their drift into the white dwarf is relatively slow. Cobbles (not shown) are too large to be dragged into the white dwarf through PR drag. 

However, when the Yarkovsky effect is activated, small boulders become susceptible to destructive drag. On the small boulder plot, we considered three different Yarkovsky efficiencies ($\xi$). The smallest boulders ($R\approx 0.1$ m) could easily be accreted for $\xi \lesssim 30\%$, whereas medium-sized boulders ($R\approx 10$ m) will not accrete, even for $\xi = 100$\%.

\subsection{Time evolution close to white dwarf}

The timescales plotted on Fig. \ref{time} are too long to appreciate the behaviour of objects close to the white dwarf. Consequently, we have provided a few zoomed-in examples for dust at two different cooling ages in Fig. \ref{cool}. Here, $t=0$ corresponds to cooling ages of 10 Myr (left panel) and 1 Gyr (right panel). Both objects in each panel are on initially circular orbits at $10R_{\odot}$, and the solid and dashed lines indicated different densities.

At such a close initial distance, the timescale to reach a typical Roche radius or sublimation distance (Fig. \ref{DesPlot}) is about $10^{2}-10^{3}$ yr for a cooling age of 10 Myr, and is about $10^4-10^5$ yr for a cooling age of about 1 Gyr. Hence, even for a relatively young white dwarf planetary system such as WD~J0914+1914 \citep{ganetal2019}, with a cooling age of about 13 Myr and an extended disc \citep{zotetal2020}, the timescale for PR drag to move an object from $10R_{\odot}$ to $1R_{\odot}$ is, at least, one human lifetime. Perhaps more promising near-term observational opportunities would arise from the movement of material from a smaller initial separation of $1R_{\odot}$ to $0.5R_{\odot}$, although cooler (and dimmer) white dwarfs would then be required for the sublimation distance to be sufficiently small.

\begin{figure*}
\centerline{\LARGE \underline{Planet-less system pollutants: maximum reach}}
\centerline{}
\centerline{
\includegraphics[width=8.5cm]{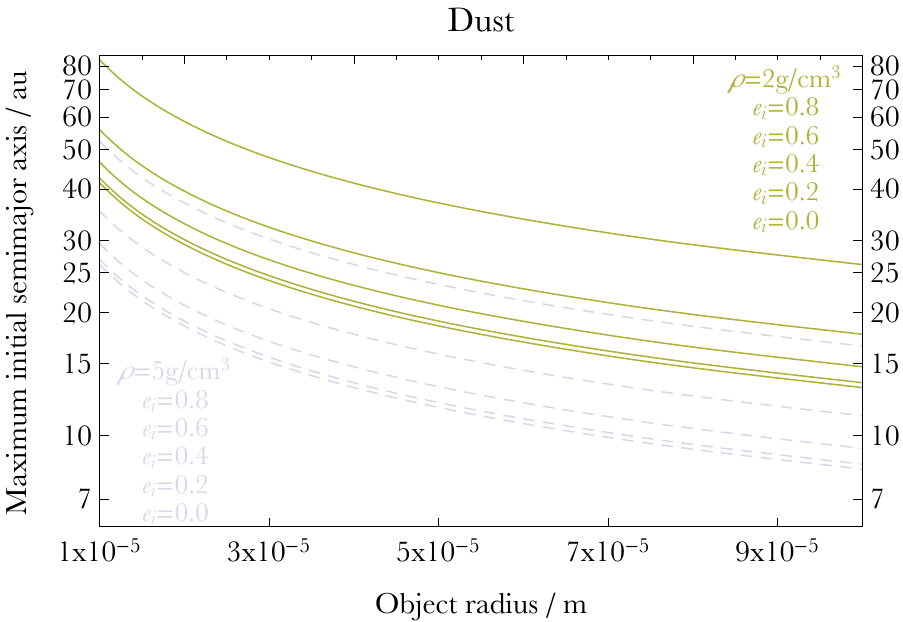}
\includegraphics[width=8.5cm]{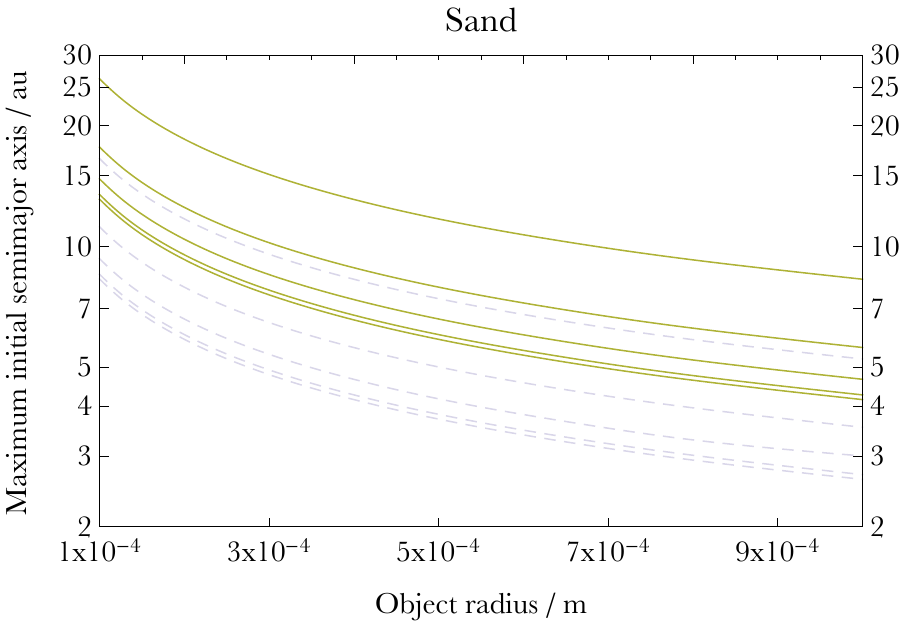}
}
\centerline{}
\centerline{
\includegraphics[width=8.5cm]{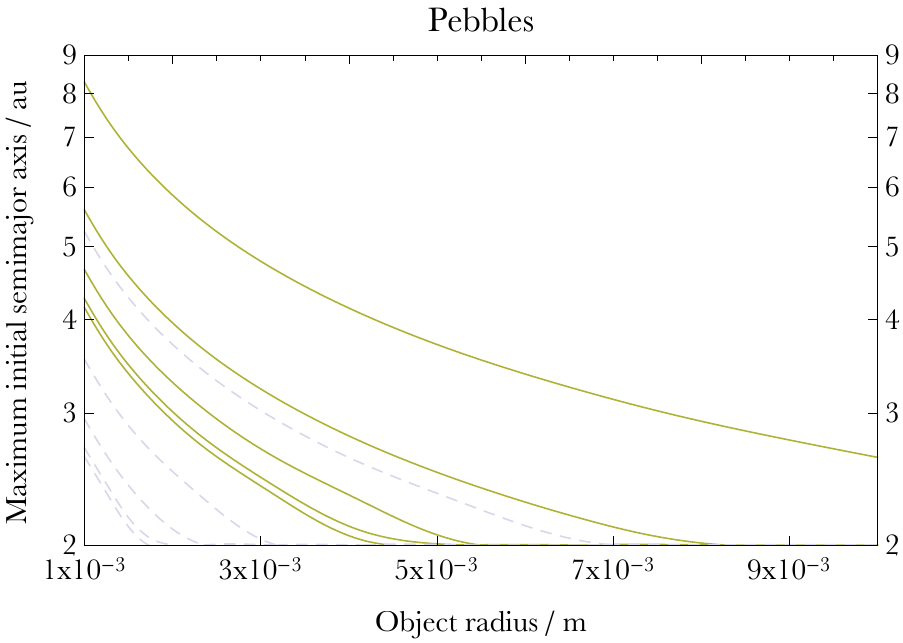}
\includegraphics[width=8.5cm]{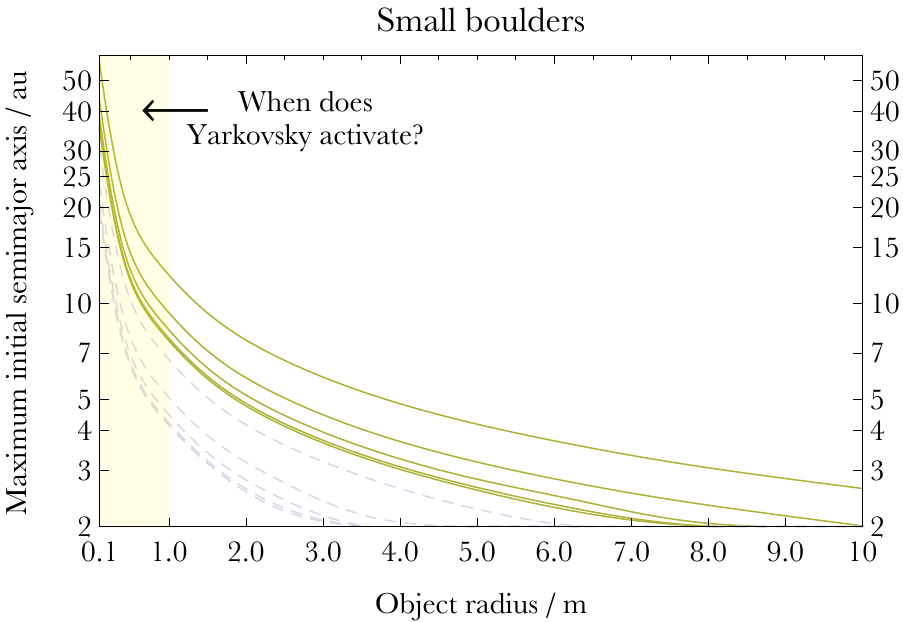}
}
\caption{
The maximum $a_{\rm i}$ ($y$-axis) values for which objects of a given radius $R$ reach a distance of $0.05R_{\odot}$ over 10 Gyr of cooling. The five green solid curves correspond to, in visually descending order, $e_{\rm i} = 0.8,0.6,0.4,0.2,0.0$, for $\rho = 2$ g/cm$^3$. The five grey dashed curves are similar, except for $\rho = 5$ g/cm$^3$. These plots quantify the parameter space in which dust, sand, pebbles and small boulders can pollute the white dwarf without the help from a planet. The plots also suggest that objects with radii in between that of pebbles and small boulders (cobbles), as well as objects larger than small boulders (asteroids), cannot pollute the white dwarf without help from a planet. 
}
\label{crit}
\end{figure*}

\subsection{Maximum reach of pollutants}

The time evolutions in Fig. \ref{time} suggest that there exist maximum values of $a_{\rm i}$ and $e_{\rm i}$ for which a given type of object could reach the close vicinity of the white dwarf within 10 Gyr of cooling. Identifying these values for given $R$ and $\rho$ required performing an ensemble of simulations and then picking out, post-simulation, the corresponding values of $a_{\rm i}$ and $e_{\rm i}$. The outcome of this exercise is only marginally dependent on the separation at which a simulation was stopped, whether that be $1.0 R_{\odot}$ or a smaller value, as shown in Fig. \ref{cool}. We chose $0.05R_{\odot}$, which corresponds to the smallest Roche radius shown in Fig. \ref{DesPlot}.

We plot these maximum values of $a_{\rm i}$ and $e_{\rm i}$ in Fig. \ref{crit}. Comparing the extent of the $y$-axes for the dust, sand and pebble plots illustrates how rapidly max($a_{\rm i}$) decreases with $R$, and why cobbles cannot pollute the white dwarf without help from a planet. Also notable is how high max($a_{\rm i}$) can reach for dust: in the most extreme case, dust which would initially reside beyond the orbit of Neptune could eventually pollute the white dwarf without any planetary perturbations. 

For small boulders, the situation is highly dependent on the minimum $R$ for which the Yarkovsky effect activates. Further, the small boulder plot in Fig. \ref{crit} assumes 100\% efficiency ($\xi = 100\%$) for each curve. Hence, this plot should be considered with these aspects in mind. Although theoretically boulders which initially reside at tens of au could be dragged close to the white dwarf, a more realistic maximum value of $a_{\rm i}$ perhaps is on the order of au.

\begin{figure*}
\centerline{\LARGE \underline{Spinning apart initial non-pollutants}}
\centerline{}
\centerline{
\includegraphics[width=8.5cm]{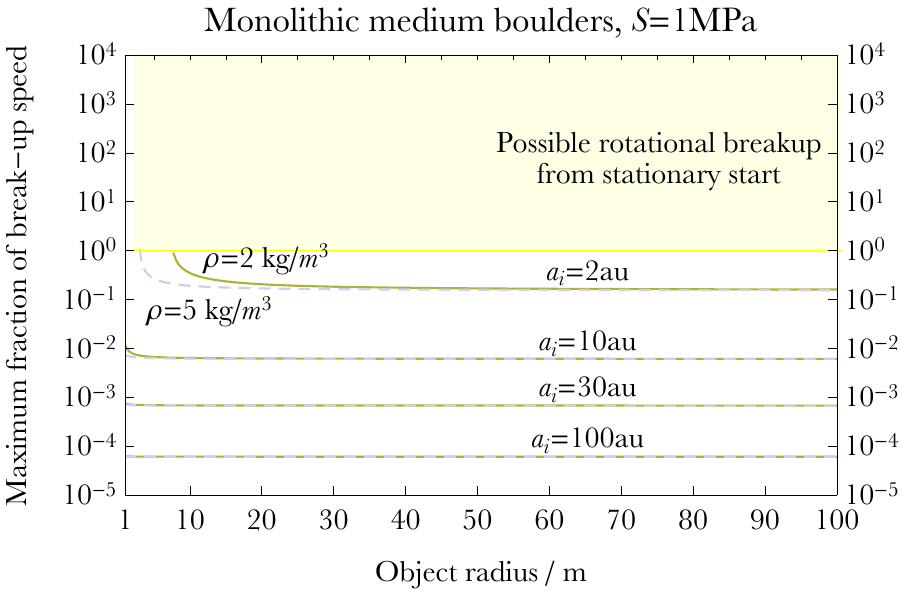}
\includegraphics[width=8.5cm]{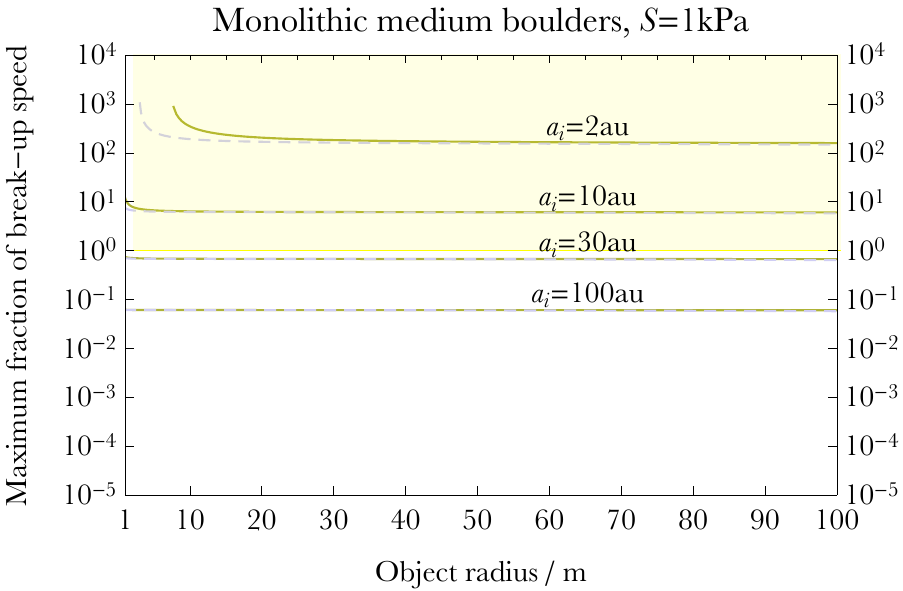}
}
\centerline{}
\centerline{
\includegraphics[width=8.5cm]{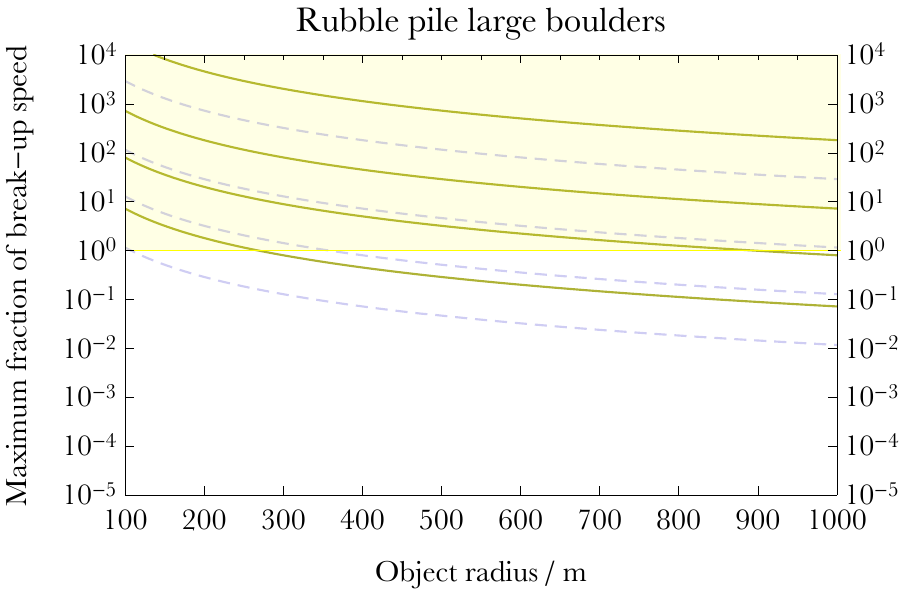}
\includegraphics[width=8.5cm]{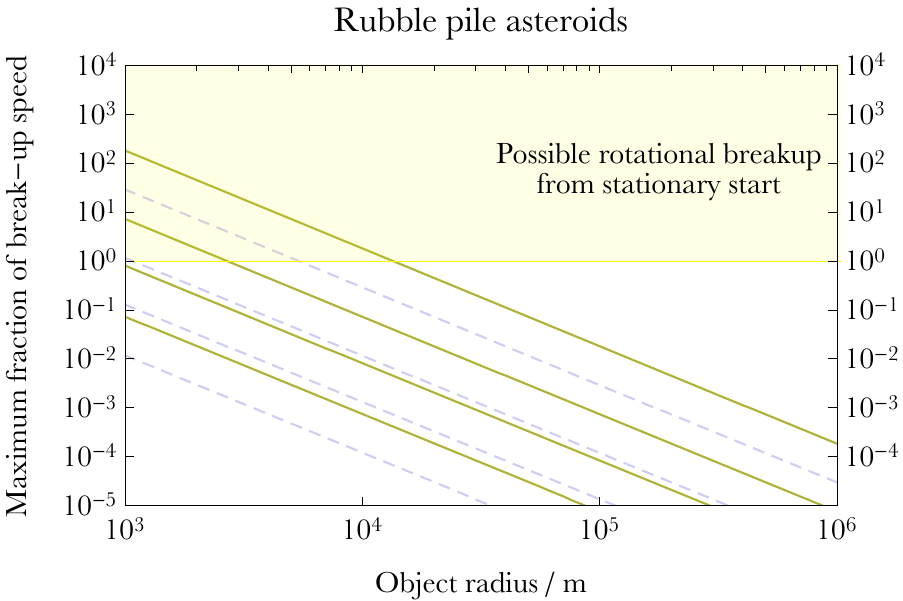}
}
\caption{
Maximum fraction of break up speed achieved due to the YORP effect. When this value is less than unity, then the YORP effect cannot spin up an initially stationary object to destruction. Further, a value greater than unity does not guarantee break-up, because the YORP effect can both speed up and slow down objects. The solid lines correspond to $\rho=2$ g/cm$^3$, and the dashed lines correspond to $\rho=5$ g/cm$^3$. The plots demonstrate that weak boulders are highly susceptible to breakup, despite the rapidly dimming white dwarf luminosity.
}
\label{spin}
\end{figure*}

\subsection{Spinning asteroids to destruction}

Although white dwarfs cannot drag asteroids to destruction without the help of planets, white dwarf luminosity does spin the asteroids up through the YORP effect. If the asteroid (or dwarf planet) is already spinning quickly, then this extra nudge could be sufficient to break it apart. The resulting fragments may then be small enough for PR drag and/or the Yarkovsky effect to take over. 

Given this possibility, here we quantify the maximum possible additional spin that could be imparted onto an object. The objects upon which the YORP effect acts are boulders and asteroids (Fig. \ref{Scales}), most of which cannot be dragged to the close vicinity of the white dwarf. Only medium-sized boulders ($R=10^{0}-10^{1}$) can both be affected by the YORP effect and can be dragged into the white dwarf through the Yarkovsky effect. 

We can compute the maximum additional spin generated while being agnostic about the initial spin of the object because of the form of Eq. (\ref{YORP}). Further, this added spin is scalable with the factor $\chi/\rho R^2$; in our simulations, we fix $\chi = 10^{-3}$ and consider two values of $\rho$ (2, 5 g/cm$^3$). If the maximum additional spin generated is less than $\omega_{\rm breakup}$ (Eq. \ref{breakup}), then an object with $\omega_{\rm i} = 0$ rad/s remains intact. Further, if the maximum additional spin is greater than $\omega_{\rm breakup}$, then the object is not guaranteed to break up because it might spin down at points during its evolution.

We display the results in Fig. \ref{spin}. The small region where the curves cut off in the upper panels indicates radii where the Yarkovsky effect drags the object to the close vicinity of the white dwarf. Hence, destruction from the Yarkovsky and YORP effects is naturally separated across almost the entire parameter space.

The figure also illustrates that for both monolithic boulders with a high internal strength of $S=1$ MPa, and for rubble pile asteroids ($S=0$ Pa), white dwarf luminosity is usually too weak to generate enough spin from a stationary start to achieve breakup. However, for weaker boulders, the total added spin which is generated could easily break up the objects. Then, depending on the size of the fragments and the cooling age at which breakup occurs, the boulder fragments may end up reaching the close vicinity of the white dwarf.

\section{Constraining initial conditions}

The formulation presented here deliberately allows one to adopt a wide range of initial conditions for planetary debris on the white dwarf phase. In this section, we discuss why the range of realistic initial conditions is so broad.

Although imaged protoplanetary discs range in size, a typical outer boundary is about 100 au \citep{andrews2020}. The remnant debris disc which survives until the end of the main sequence then expands by a factor of 2-3 due to stellar mass loss, even amidst a variety of radiative and gravitational forces which reshape the contents. These other forces include mutual collisions, stellar wind pressure, stellar wind drag, PR drag, the Yarkovsky effect and the YORP effect. 

\cite{bonwya2010} modelled the evolution of different combinations of the first four of these forces in debris discs with no planets across the giant branch phases of evolution. They hence obtained a range of values for the maximum diameters at which objects can be removed from the system during this evolution: $10^{-1}$--$10^3$ m depending on distance, from 6 to about 300 au. However, they did not consider the Yarkovsky and YORP effects.

Building on the work of \cite{veretal2014b}, \cite{versch2020} modelled (with no planets) a type of YORP cascade, where increasing stellar luminosity can produce successive break-ups for boulders and asteroids larger than about $10^0$ m. This type of cascade is most effectual near the tip of the asymptotic giant branch (right before the star becomes a white dwarf), and allows objects ranging in size from dust to boulders to be freshly generated immediately before the white dwarf phase. They demonstrated that breakup could occur out to distances of hundreds of au when $R \lesssim 10^2$ m.

Further, boulders which survive all these processes are subject to either inward or outward migration from the Yarkovsky effect. In our computations here, we used only a limiting value of the Yarkovsky effect in order to place bounds on boulder motion around white dwarfs. However, as evidenced by V-shaped asteroid family signatures in the solar system, both inward and outward drift are about equally as common, at least for a static $1L_{\odot}$ luminosity star \citep[e.g.][]{boletal2017}. Along the giant branches, these drifts are ``supercharged" due to the higher and time-variable stellar luminosity, and can be on the scale of tens or hundreds of au \citep{veretal2015a,veretal2019a}. Such fast drifts can then lead to collisional evolution outside of the standard framework as modelled by, e.g. \cite{bonwya2010}, particularly when at least one planet remains \citep{zotver2020}.

\

\section{Potentially testable predictions}

Ideally, a pollution scenario without planets can be observationally distinguished from one where at least one planet is present. The reality, however, is that achieving such a distinction is not yet possible because of degeneracies and physical unknowns. In this section, we quantitatively address these issues. 

There are at least two potential relevant observables: (1) the total mass of dust which may be detectable beyond the Roche sphere, and (2) the accretion rate onto the white dwarf photosphere. The first observable would likely represent the better distinguisher of the pollution scenarios because fewer assumptions about that matter's dynamical and physical history would be required.

Quantitatively, for warm white dwarfs, solid masses as low as $10^{15}$ kg may be detectable at distances of several solar radii \citep{bonetal2017}. Accretion rates onto the photosphere typically lie in the range $10^5-10^{11}$ g/s ($10^2-10^8$ kg/s) \citep{giretal2012,beretal2014} and could show a pronounced decrease with cooling age \citep{holetal2018} or remain relatively flat as a function of time \citep{bloxu2021}. These values represent tangible numbers on which a comparison with theories may be achieved.

Whether the matter is delivered quickly or slowly to the white dwarf depends on the presence or lack of planets in the system. Quick delivery is due to gravitational perturbations from a planet, such that an asteroid breaks up at or inside of the Roche sphere \citep{debetal2012,veretal2014a,malper2020b,broetal2021,lietal2021}, or outside of the Roche sphere but within about $3r_{\rm Roche}$ \citep{veretal2020}. Alternatively, one method of slow delivery is quantified here, where the polluting matter is dragged to the white dwarf without the aid of a planet. 

Hence, the distribution of matter at the end of giant branch phases is key. Let us first consider a single asteroid of mass $M_{\rm ast}$ that breaks up due to the YORP effect. Assume that the asteroid is composed of constituent particles that follow a smooth power law with exponent $p$ such that the asteroid's size distribution can be described as

\

\begin{equation}
\frac{dN}{dM} = u M^{-p},
\label{power}
\end{equation} 

\noindent{}where $N$ represents the number of constituent bodies. It follows that

\begin{equation}
u = \frac{\left(p-2\right)M_{\rm ast}}{M_{\rm min}^{2-p} - M_{\rm max}^{2-p}},
\end{equation}

\noindent{}where $M_{\rm min}$ and $M_{\rm max}$ represent the least massive and most massive constituents. 

We must discretise Eq. (\ref{power}) in order to apply it to Eqs. (\ref{PRa})-(\ref{Yarke}) because the differential equations do not admit, to our knowledge, an analytical solution. In this respect, we note that the number of constituent particles with masses in-between $M$ and $M + \Delta M$ equals $u M^{-p} \Delta M$. 

We then assume that all constituent particles are strengthless ($S=0$) spheres and have a common density $\rho$, which we will further assume equals 2 g/cm$^3$. Doing so allows us to convert $u M^{-p} \Delta M$ into expressions involving constituent radii $R$ for application into our differential equations. Assuming $R_{\rm min} = 10^{-5}$ m and $R_{\rm max} = 10^{1}$ m allows us to sample the evolution of dust, sand, pebbles, cobbles and small boulders.

The resolution at which different $R$ values are sampled are limited by computational power. We sample 20 different $R$ values per decade in logarithmic space assuming $M_{\star} = 0.6M_{\odot}$, $\xi = 1$ and an initially $a_{\rm i}=10$ au orbit with a non-zero but also non-negligible initial orbital eccentricity of $e_{\rm i}=0.1$. We choose $p=3.0$ simply because that value mirrors the distribution on the surface of Itokawa, without having access to its interior \citep{micetal2008}.

The results of this computation confirm that cobbles and the larger boulders never reach the white dwarf, the smallest dust accretes onto the white dwarf almost immediately, and the remainder of asteroid constituents reach the white dwarf at different cooling times. When applicable, we track the times at which a constituent reaches the following thresholds in distance: $10R_{\odot}$, $5R_{\odot}$, $2R_{\odot}$, and $1R_{\odot}$. 

Because of our limited resolution in initial conditions, and the inability for us to analytically invert the equations of motion as a function of cooling age, let us consider a specific instance. We find that sand which was sampled with $R=1.679 \times 10^{-3}$ m reaches $10R_{\odot}$, $5R_{\odot}$, $2R_{\odot}$, and $1R_{\odot}$ at cooling ages of $2.92053$, $2.92196$, $2.92236$, and $2.92242$ Gyr, respectively. The number of particles of this sand is about $3.11 \times 10^{22}$, amounting to a total mass of $1.2 \times 10^{-4}M_{\rm ast}$.

Hence, from this one asteroid of mass $M_{\rm ast}$, about $10^{-4}M_{\rm ast}$ of it would reach a distance of $10R_{\odot}$ almost 3~Gyr into white dwarf cooling. This mass then takes another 1.43 Myr to reach $5R_{\odot}$, and then another 0.4 Myr to reach $2R_{\odot}$, and finally another 0.06~Myr to reach $1R_{\odot}$. This drift rate would be quicker for hotter and younger white dwarfs, and slower for older and cooler white dwarfs.

Additionally, regarding the mass budget, if one assumes that this asteroid had $10^4$ physically identical companions all on the same orbits which broke up in exactly the same way, and assumes no other extant particles in the system, then a total mass of $M_{\rm ast}$ would arrive in the vicinity of the white dwarf at this cooling age. More realistically, at this cooling age there would be contributions from differently-sized constituent particles originating from different asteroids with different physical properties and initial orbits. In this respect, the level of degeneracy is very high. 

Adding to the degeneracies are two potentially important extra considerations. The first is that gravitational, erosional, collisional and viscous evolution amongst constituents can occur even while en route to the white dwarf, producing a change in disc structure as well as the number of disc constituents. \cite{verhen2020} computed the lifetimes of these particulate debris discs on different spatial scales around white dwarfs by incorporating these physical effects together. They found that, depending on the structure and physical properties of the disc, on scales of au these disc lifetimes range widely from $10^0 - 10^{10}$ yr, and on scales of tens of au, the bounds on this rough interval increase by several orders of magnitude. 

The second extra source of degeneracy is that at around $1r_{\rm Roche}$, gas is produced as some particles sublimate (see Fig. \ref{DesPlot}) and the gas-dust mixture generates additional physical processes. The lifetimes of the discs in this close vicinity to the white dwarf have been observationally empirically estimated to be on the order of $10^4-10^6$ yr \citep{giretal2012}, but the uncertainty in these estimates is significant because these estimates were generated by combining constraints from two different classes of white dwarfs. Subsequent accretion onto the white dwarf photosphere could then be in a leading, steady-state or trailing phase \citep{koester2009}, not even accounting for external replenishment.

Overall, we believe that the best way to distinguish pollution scenarios is to observe dust beyond about $3r_{\rm Roche}$, beyond which tidal or chaotic rotational disruption has not been shown to be possible (whereas YORP break-up is; Fig. \ref{spin}). Being able to place constraints on the total dust mass in distance intervals on the order of $r_{\rm Roche}$ would further help break the degeneracy in the parameter space (e.g. the distributions of $M_{\rm ast}$, $M_{\rm min}$, $M_{\rm max}$, $p$, $a_{\rm i}$, $e_{\rm i}$, $S$, $\rho$, $\xi$, and the evolution properties of the au-scale discs) of the origin scenarios.

\section{Summary}

We have shown that white dwarf radiation is capable of dragging dust, sand, pebbles and small boulders from au-scale distances into the star's immediate vicinity without the help of a planet. In these systems the planets were already lost before our simulations began. The maximum separations for which this inward drag is possible (Fig. \ref{crit}) can be on the scale of au or tens of au depending primarily on object radius, density, and initial eccentricity. Further, white dwarf radiation is strong enough to spin up large boulders to destruction (Fig. \ref{spin}), allowing the smaller fragments to be dragged towards the star. Overall, although white dwarfs are now known to host planets, such planets are not a required component of every polluted system.

\section*{Acknowledgements}

DV gratefully acknowledges the support of the STFC via an Ernest Rutherford Fellowship (grant ST/P003850/1). 

\section*{Data Availability}

The simulation inputs and results discussed in this paper are available upon reasonable request to the corresponding author.

\label{lastpage}
\end{document}